\documentclass[preprint,showpacs,preprintnumbers,amsmath,amssymb]{revtex4}
\usepackage{booktabs}
\usepackage{mathrsfs}
\usepackage{epsfig}
\usepackage{graphicx}
\usepackage{dcolumn}
\usepackage{bm}
\usepackage{amsmath}
\usepackage{slashed}       

\let\jnfont=\rm
\def\NPB#1,{{\jnfont Nucl.\ Phys.\ B }{\bf #1},}
\def\PLB#1,{{\jnfont Phys.\ Lett.\ B }{\bf #1},}
\def\EPJC#1,{{\jnfont Eur.\ Phys.\ Jour.\ C }{\bf #1},}
\def\PRD#1,{{\jnfont Phys.\ Rev.\ D }{\bf #1},}
\def\PRL#1,{{\jnfont Phys.\ Rev.\ Lett.\ }{\bf #1},}
\def\MPLA#1,{{\jnfont Mod.\ Phys.\ Lett.\ A }{\bf #1},}
\def\JPG#1,{{\jnfont J.\ Phys.\ G}{\bf #1},}
\def\CTP#1,{{\jnfont Commun.\ Theor.\ Phys.\ }{\bf #1},}
\def\ZPC#1,{{\jnfont Z.\ Phys.\ C }{\bf #1},}
\def\JHEP#1,{{\jnfont JHEP \ }{\bf #1},}
\def\lsim{\raise0.3ex\hbox{$<$\kern-0.75em\raise-1.1ex\hbox{$\sim$}}}
\def\gsim{\raise0.3ex\hbox{$>$\kern-0.75em\raise-1.1ex\hbox{$\sim$}}}

\begin{document}

\title{ Di-photon Higgs signal at the LHC: a comparative study in
        different supersymmetric models}

\author{Junjie Cao$^1$, Zhaoxia Heng$^1$, Tao Liu$^2$, Jin Min Yang$^2$}

\affiliation{
  $^1$  Department of Physics,
       Henan Normal University, Xinxiang 453007, China \\
  $^2$ Key Laboratory of Frontiers in Theoretical Physics,
      Institute of Theoretical Physics, Academia Sinica, Beijing 100190, China
      \vspace{1cm}}

\begin{abstract}
As the most important discovery channel for a light Higgs boson at
the LHC, the di-photon signal $gg\to h\to \gamma\gamma$ is sensitive
to underlying physics. In this work we investigate such a signal in a
comparative way by considering three different supersymmetric models,
namely the minimal supersymmetric standard model (MSSM), the
next-to-minimal supersymmetric standard model (NMSSM) and the nearly
minimal supersymmetric standard model (nMSSM).
Under the current collider and cosmological constraints
we scan over the parameter space and obtain the following
observation in the allowed parameter space:
(i) In the nMSSM the signal rate is always suppressed;
(ii) In the MSSM the signal rate is suppressed in most cases, but in
a tiny corner of the parameter space it can be enhanced (maximally
by a factor of 2);
(iii) In the NMSSM the signal rate can be enhanced or  suppressed
depending on the parameter space, and the enhancement factor
can be as large as 7.
\end{abstract}

\maketitle

\section{Introduction}
So far the most important question to be answered in particle
physics is the mechanism of the electroweak symmetry breaking and
thus hunting for the Higgs boson responsible for it is the main task
of current collider experiments.  In the framework of the Standard
Model (SM), the mass of the Higgs boson is preferred to be
$116.4^{+15.6}_{-1.3}$ GeV by precision electroweak data
\cite{h-mass-ew}. To search for such a relatively light Higgs boson,
great efforts have been made in LEP and Tevatron experiments, which
reported null results and excluded a Higgs boson with $m_h \leq
114.4{\rm ~GeV}$ \cite{Barate:2003sz} and $158{\rm ~GeV} \leq m_h
\leq 175 {\rm ~GeV}$ \cite{Tevatron-2010} at 95$\%$ C.L.. The Large
Hadron Collider (LHC) is more powerful in discovering the SM Higgs
boson, and depending on its mass, different search strategies will
be applied. For a light Higgs boson below 140 GeV, although its
largest signal at the LHC is $b\bar{b}$ from the gluon-fusion
process $gg\to h \to b \bar{b}$ \cite{LHCHiggsCross}, such a signal
is undetectable due to the overwhelming QCD background; instead, the
rare decay mode $h\to \gamma\gamma$ with $Br (h \to \gamma \gamma)
\simeq 0.2\%$ for $m_h = $120 GeV offers a very clean signature to
make the di-photon signal $gg\to h\to \gamma\gamma$ a promising
discovery channel. It is now expected that, with $2 fb^{-1}$
integrated luminosity at the LHC running at $\sqrt{s}=7$ TeV, the
di-photon signal is able to exclude the light Higgs boson in the SM
\cite{LHC}.

In low energy supersymmetric models (SUSY), the SM-like Higgs boson
(the CP-even Higgs boson with largest coupling to gauge bosons) is
usually predicted with mass below about 140 GeV. For such a Higgs
boson, although there may exist other discovery channels at the LHC,
the di-photon channel $gg\to h\to \gamma\gamma$ is still one of the
most important discovery modes. So, studying this signal will allow
for a probe of low energy SUSY and,  as emphasized in \cite{Yuan},
even a discrimination of different models. Although in the
literature some studies of the signal have been presented in SUSY
\cite{diphoton1,diphoton2,diphoton3}, these analyses were performed
separately in different models and a comparative study is necessary
in order to discriminate the models. In this work we perform such a
comparative study by considering three different SUSY models, namely
the minimal supersymmetric standard model (MSSM), the
next-to-minimal supersymmetric standard model (NMSSM)
\cite{NMSSM1,NMSSM2} and the nearly minimal supersymmetric standard
model (nMSSM) \cite{xnMSSM,cao-xnmssm}. We will scan over the
parameter space under current constraints from collider experiments
and the neutralino dark matter relic density, and then in the
allowed parameter space we calculate the di-photon signal rate and
compare the results for different models.

This work is organized as follows. We first briefly describe the
three supersymmetric models in Sec. II. Then we present our
numerical results and discussions in Sec. III. Finally, we draw our
conclusions in Sec. IV.

\section{Supersymmetric Models}
As the most economical realization of SUSY in particle physics, the
MSSM has been intensively studied. However, since this model suffers
from the $\mu-$problem and the little hierarchy problem, some of its
extensions like the NMSSM and nMSSM were recently paid attention
to\cite{NMSSM1}. The differences of these models come from their
superpotentials:
\begin{eqnarray}
 W_{\rm MSSM}&=& W_F + \mu \hat{H_u}\cdot \hat{H_d}, \label{MSSM-pot}\\
 W_{\rm NMSSM}&=&W_F + \lambda\hat{H_u} \cdot \hat{H_d} \hat{S}
 + \frac{1}{3}\kappa \hat{S^3},\\
 W_{\rm nMSSM}&=&W_F + \lambda\hat{H_u} \cdot \hat{H_d} \hat{S}
 +\xi_FM_n^2\hat S  \label{nMSSM-pot} ,
\end{eqnarray}
where $W_F$ is the MSSM superpotential without the $\mu$ term,
$\hat{H}_{u,d}$ and $\hat S$ are the Higgs doublet and singlet
superfields respectively, and the dimensionless coefficients
$\lambda$, $\kappa$ and $\xi_F$ and the dimensional coefficients
$\mu$ and $M_n$ are usually treated as independent parameters.  In
the NMSSM and nMSSM, when the scalar component ($S$) of the singlet
Higgs superfield $\hat{S}$ develops a vacuum expectation value
(VEV), an desired effective $\mu$-term ($\mu_{eff}= \lambda \langle
S \rangle$) is generated at the weak scale. Note that the nMSSM
differs from the NMSSM in the last term of the superpotential, where
the cubic singlet term $\kappa \hat{S^3}$ in the NMSSM is replaced
by the tadpole term $\xi_FM_n^2$. Considering that the tadpole term
does not induce any interaction, one can infer that, except for the
minimization conditions of the Higgs potential and the mass matrices
of the Higgs bosons, the nMSSM is actually identical to the NMSSM
with vanishing $\kappa$.

Corresponding to Eq.(\ref{MSSM-pot}-\ref{nMSSM-pot}), the
soft-breaking terms in Higgs sector are given by
\begin{eqnarray}
V_{\rm soft}^{\rm MSSM}&=&\tilde m_u^2|H_u|^2 + \tilde m_d^2|H_d|^2
+ (B\mu H_u\cdot H_d + h.c.),\\
V_{\rm soft}^{\rm NMSSM}&=&\tilde m_u^2|H_u|^2 + \tilde m_d^2|H_d|^2
+ \tilde m_S^2|S|^2 +(A_\lambda \lambda SH_u\cdot H_d
+\frac{A_\kappa}{3}\kappa S^3 + h.c.),\\
V_{\rm soft}^{\rm nMSSM}&=& \tilde m_u^2|H_u|^2 + \tilde
m_d^2|H_d|^2 + \tilde m_S^2|S|^2 +(A_\lambda \lambda SH_u\cdot H_d
+\xi_S M_n^3 S + h.c.),
\end{eqnarray}
where $\tilde{m}_{u}$, $\tilde{m}_{d}$, $\tilde{m}_{S}$, $B$,
$A_\lambda$ and $A_\kappa$ are all soft parameters. Like the usual
treatment of the multiple Higgs theory, one can write the scalar
fields $H_u$, $H_d$ and $S$ as
\begin{eqnarray}
H_u = \left ( \begin{array}{c} H_u^+ \\
       \frac{v_u + \phi_u + i \varphi_u}{\sqrt{2}}
        \end{array} \right),~~
H_d & =& \left ( \begin{array}{c}
             \frac{v_d + \phi_d + i \varphi_d}{\sqrt{2}}\\
             H_d^- \end{array} \right),~~
S  = \frac{1}{\sqrt{2}} \left( s + \sigma + i \xi \right),
\end{eqnarray}
and diagonalize the mass matrices of the Higgs bosons to get their
mass eigenstates:
\begin{eqnarray} \left( \begin{array}{c} h_1 \\
h_2 \\ h_3 \end{array} \right) = U^H \left( \begin{array}{c} \phi_u
\\ \phi_d\\ \sigma\end{array} \right),~ \left(\begin{array}{c} a\\
A\\ G^0 \end{array} \right) = U^A \left(\begin{array}{c} \varphi_u
\\ \varphi_d \\ \xi \end{array} \right),~ \left(\begin{array}{c} H^+
\\G^+ \end{array}  \right) =U \left(\begin{array}{c}H_u^+\\ H_d^+
\end{array} \right).  \label{rotation}
\end{eqnarray}
In above expressions, $h_1,h_2,h_3$ and $a,A$ denote physical
 CP-even and CP-odd neutral Higgs bosons respectively,  $G^0$ and
$G^+$ are Goldstone bosons eaten by $Z$ and $W^+$, and $H^+$ is the
charged Higgs boson. Note in the MSSM,  due to the absence of $S$
there only exist two CP-even Higgs bosons and one CP-odd Higgs
boson, and consequently, $U^H$ and $U^A$ are reduced to $2 \times 2
$ matrices parameterized by the mixing angles $\alpha$ and $\beta$
respectively. In our study, we choose the input parameters in the
Higgs sector as ($\tan\beta$, $m_A$, $\mu$) for the MSSM,
($\lambda$, $\kappa$, $\tan\beta$, $\mu_{eff}$, $m_A$, $A_\kappa$)
for the NMSSM with $m_A^2 = \frac{2\mu}{\sin 2 \beta} (A_\lambda +
\frac{\kappa\mu}{\lambda})$, and ($\lambda$, $\tan \beta$,
$\mu_{eff}$, $A_\lambda$, $\tilde{m}_S$, $m_A$) for the nMSSM with
$m_A^2 = \frac{2}{\sin 2 \beta} ( \mu A_\lambda + \lambda \xi_F
M_n^2$).

The Yukawa couplings of the neutral Higgs bosons to the top and
bottom quarks are given by \cite{NMSSM1}
\begin{eqnarray}
{\cal L}_{\rm Yukawa}&=&-\frac{gm_t}{2m_W\sin\beta}U^H_{i1}\bar t t h_i
-\frac{gm_b}{2m_W\cos\beta}U^H_{i2}\bar b b h_i\nonumber\\
&&+\frac{igm_t}{2m_W\sin\beta}U^A_{11}\bar t\gamma_5 t a
+\frac{igm_b}{2m_W\cos\beta}U^A_{12}\bar b\gamma_5 b a,
\label{Yukawa}
\end{eqnarray}
with $U^H$, $U^A$ defined in Eq.(\ref{rotation}). Obviously, once
$U^H_{i2}/\cos\beta \ll 1$ as discussed later, the width of $h_i \to
b \bar{b}$ is to be suppressed.

Note the properties of the lightest neutralino $\tilde{\chi}_1^0$ in
the nMSSM are quite peculiar \cite{cao-xnmssm}. After diagonalizing
the neutralino mass matrix in the nMSSM, its mass takes the form
\cite{lsp-mass}
\begin{eqnarray}
m_{\tilde{\chi}^0_1} \simeq \frac{2\mu \lambda^2 (v_u^2+
v_d^2)}{2\mu^2+\lambda^2  (v_u^2+ v_d^2)}
             \frac{\tan \beta}{\tan^2 \beta+1},
             \label{mass-exp}
\end{eqnarray}
which implies that $\tilde{\chi}_1^0$ must be lighter than about
$60$ GeV for $\mu > 100 {\rm ~GeV}$ (required by chargino mass bound)
and $\lambda < 0.7$ (required by perturbativity). If
$\tilde{\chi}_1^0$ acts as the dark matter candidate, a light CP-odd
Higgs boson $a$ is then preferred to accelerate $\tilde{\chi}_1^0$
annihilation to get the acceptable dark matter relic
density\cite{cao-xnmssm}. Detailed study indicates that
$m_{\tilde{\chi}^0_1} \leq 37 {\rm ~GeV}$ and for most cases, $m_a
\leq 60 {\rm ~GeV}$, which implies the SM-like Higgs boson $h$ may
decay into $\tilde{\chi}_1^0 \tilde{\chi}_i^0$ or $aa$ so that $Br
(h \to \gamma \gamma) $ is suppressed\cite{cao-xnmssm}.

\section{Numerical Results and Discussions}
\subsection{Description of calculations}
\label{sec-a}
To compare the signal rate with the SM prediction, we define
a normalized rate as
\begin{eqnarray}
R^{\rm SUSY} &\equiv & \sigma_{SUSY} ( p p \to h \to
\gamma \gamma)/\sigma_{SM} ( p p \to h \to \gamma \gamma ) \nonumber \\
&\simeq& [\Gamma(h\to gg) Br(h\to \gamma\gamma)] /[\Gamma(h_{SM}\to
gg) Br(h_{SM}\to \gamma\gamma)] \nonumber \\
&=&  [\Gamma(h\to gg) \Gamma(h\to \gamma\gamma)] /[\Gamma(h_{SM}\to
gg) \Gamma(h_{SM}\to \gamma\gamma)] \times
\Gamma_{tot}(h_{SM})/\Gamma_{tot}(h)  \label{definition}
\end{eqnarray}
where we used the narrow width approximation and the fact that at
leading order the cross section of the parton process $g g \to h$
is correlated with the decay width of $h \to g g$ by
\begin{eqnarray}
\hat \sigma (gg\to h) = \sigma_0^h m^2_h\delta(\hat s-m_h^2)
 = \frac{\pi^2}{8m_h}\Gamma (h\to gg)\delta(\hat s-m_h^2).
\end{eqnarray}
In SUSY, the $h\gamma\gamma$ coupling arises mainly from the loops
mediated by W-boson, charged Higgs boson, charginos and the third
generation fermions and sfermions, and the $h g g$ coupling only
from the loops mediated by third generation quarks and squarks.
Consequently, the widths of $h \to \gamma \gamma, g g$ are given by
\cite{diphoton1}
\begin{eqnarray}
 \Gamma(h\to \gamma\gamma)&=&\frac{G_{F}\alpha^{2}m_{h}^{3}}{128\sqrt{2}\pi}
\left| \sum_f N_{c}\, Q_{f}^{2}\, g_{hff}\, A_{1/2}^{h}(\tau_{f})+
g_{hWW}\, A_{1}^{h}(\tau_{W}) + {\cal A}^{\gamma\gamma}\right|^2, \label{hgaga}\\
 \Gamma(h\to gg)&=&\frac{G_F \alpha_s^2 m_h^3}{36 \sqrt{2}\pi^3}
       \left| \sum_q N_{c}\, Q_{q}^{2}\,  g_{hqq}\, A_{1/2}^h(\tau_q) +  {\cal A}^{gg}
     \right|^2 \label{hgg}
\end{eqnarray}
where $\tau_i = m_h^2/(4m_i^2)$, and
\begin{eqnarray}
{\cal A}^{\gamma\gamma} &=&
 g_{hH^{+}H^{-}}\frac{m_{W}^{2}}{m^{2}_{H^{\pm}}}A_{0}^{h}(\tau_{H^{\pm}}) +
\sum_f N_c Q_f^2 g_{h\tilde{f}\tilde{f}}\frac{m_Z^2}{m^2_{\tilde{f}}}A_0^h(\tau_{\tilde{f}})
 +\sum_i g_{h\chi_i^+\chi_i^-}\frac{m_W}{m_{\chi_i}} A_{1/2}^h(\tau_{\chi_i}) , \nonumber\\
{\cal A}^{gg} &=&  \sum_{i} N_{c}Q_{q}^{2} g_{h\tilde{q}_i\tilde{q}_i}
 \frac{m_Z^2}{m_{\tilde{q}_i}^2}A_{0}^h(\tau_{\tilde{q}_i}),
\end{eqnarray}
represent pure SUSY contributions with $m_{\tilde{f}}$ and
$m_{\chi_i}$ being sfermion mass and chargino mass respectively.
Noting the asymptotic behavior of $A_i^h$ in the limit $\tau_i \ll
1$\cite{Djouadi:1998az}
\begin{equation}\label{asymp}
A_0^h \to - \frac13\ , \quad A_{1/2}^h \to -\frac43\ ,\quad
A_{1}^h \to +7 \ ,
\end{equation}
one can easily learn that the effects of the third generation
squarks on the $h\gamma \gamma$ and $h g g$ couplings drop quickly
as the squarks becomes heavy, and that the charged Higgs
contribution to $h\gamma \gamma$ coupling is usually far smaller
than the $W$-boson contribution.

In SUSY, the third generation squarks can also affect the masses and
the couplings of the CP-even Higgs bosons by radiative corrections,
and such effects are maximized in the so-called ``maximal mixing''
($m_h^{max}$) scenario defined as $X_t=2M_{\rm SUSY}$ and $A_t=A_b$
in the on-shell scheme \cite{Carena}, where $X_t =
A_t-\mu/\tan\beta$ with $A_t$ denoting the trilinear couplings of
the top squarks and $M_{SUSY}$ standing for the common soft breaking
mass for the third generation squarks, i.e., $
M_{Q_3}=M_{U_3}=M_{D_3} = M_{\rm SUSY} $. Since the corrections are
vital for our results, we will specially discuss them  later.

Different from previous studies in
\cite{diphoton1,diphoton2,diphoton3,Yuan}, we consider more
constraints on the models, which are:
\begin{itemize}
\item[(1)] The constraints from the LEP-II direct search for neutral Higgs
     bosons in various possible channels.
\item[(2)] The direct mass bounds on sparticles and Higgs boson from LEP
     and the Tevatron experiments \cite{PDG2010}.
\item[(3)] The LEP-I constraints on invisible $Z$ decay:
     $\Gamma(Z\to  \tilde\chi_1^0 \tilde\chi_1^0) < 1.76~{\rm MeV}$,
     and the LEP-II constraints on neutralino productions
     $\sigma(e^+e^-\to \tilde\chi_1^0 \tilde\chi_i^0) < 10^{-2}~{\rm pb}~ (i>1)$
     and $\sigma(e^+e^-\to \tilde\chi_i^0 \tilde\chi_j^0) < 10^{-1}~{\rm  pb}~ (i,j>1)$
     \cite{Abdallah}.
\item[(4)] The indirect constraints from B-physics (such as $b\to s\gamma$)
     and from the precision electroweak observables
     such as $\rho_{\ell}$, $\sin^2 \theta_{eff}^{\ell}$ and
     $M_W$, or their combinations $\epsilon_i (i=1,2,3)$ \cite{Altarelli}.
     We require $\epsilon_i$ to be compatible with the
     LEP/SLD data at $95\%$ confidence level.
     We also require new physics prediction of $R_b= \Gamma (Z \to \bar{b} b) /
     \Gamma ( Z \to {\rm hadrons} )$ is within the $2 \sigma$ range of
     its experimental value. The latest results for $R_b $ are $R_b^{exp}
     = 0.21629 \pm 0.00066 $ and $R_b^{SM} = 0.21578 $ for $m_t = 173$ GeV \cite{PDG2010}.
\item[(5)] The constraints from Tevatron experiments on
$\sigma(p\bar p\to h + X \to 4 \mu, 2 \mu 2 \tau)$\cite{Abazov}.
\item[(6)] The constraints from the muon anomalous magnetic moment:
     $a_\mu^{exp} - a_\mu^{SM} = (25.5 \pm 8.0 )\times 10^{-10}$ \cite{Davier}.
     We require the SUSY effects to explain $a_{\mu}$ at $2\sigma$ level.
\item[(7)] Dark matter constraints from the WMAP relic desity
     0.0975 $< \Omega h^2 <$ 0.1213 \cite{Dunkley}. For each model we
     assume the lightest neutralino as the only component for the dark matter.
\end{itemize}
As verified by numerous studies, these constraints show strong
preference on the SUSY parameters, e.g., the constraint (1) favors
heavy top squarks with significant chiral mixing, while the
constraint (6) favors large $\tan \beta$ for moderately heavy
sleptons. Note that most of the constraints have been encoded in the
program NMSSMTools \cite{NMSSMTools}, which computes various Higgs
decay rates up to one-loop level (the
dominant one-loop and leading logarithmic two-loop corrections to
the Higgs masses and mixings are also included).
We extend the code by adding more constraints in item (4) \cite{STU}
and further make it applicable to the nMSSM \cite{cao-xnmssm} (through
some helpful discussions with the authors of the NMSSMTools).

Since the LHC is now testing the probability of the enhanced
di-photon signal, we investigate the situation where the signal rate
can exceed its SM prediction. Eq.(\ref{definition}) indicates two
mechanisms in doing this. One is to enhance the $h \gamma \gamma$
coupling or the $h g g$ coupling. However, as indicated by our
numerical results, this mechanism can only enhance the couplings by
a factor up to 1.3 and 1.1, respectively. The reason is that the
relevant SUSY parameters, such as $\tan \beta$ and the third
generation squark masses, have been limited by the constraints. The
other mechanism, which proves to be capable in enhancing $R^{SUSY}$
by a factor up to 5, is to suppress the width of $h \to b \bar{b}$
to enhance the branching ratio of $h \to \gamma\gamma$. To
understand this, let's look at the expression of $\Gamma( h_i \to b
\bar{b})$, which, after including the important SUSY correction to
bottom quark mass $\Delta_b$, is given by\cite{Carena-eff}
\begin{eqnarray}
\Gamma (h_i \to b \bar{b}) \propto \left ( \frac{U_{i2}}{\cos \beta}
\right )^2 \left ( \frac{1 + U_{i1}^H/U_{i2}^H \cot \beta
\Delta_b}{1 + \Delta_b} \right )^2 \label{width}
\end{eqnarray}
where the first factor comes from the bottom Yukawa coupling in
Eq.(\ref{Yukawa}) and the second factor arises from transforming the
Higgs fields from weak basis to mass eigenstates in the low energy
effective Lagrangian. Obviously, once $U_{i2}/\cos \beta \ll 1$
and/or $U_{i1}^H/U_{i2}^H \cot \beta \Delta_b \to -1 $, $\Gamma( h_i
\to b \bar{b})$ will be greatly suppressed. In the following, we
take the MSSM as an example to discuss how to satisfy the
conditions.

In the MSSM, Eq.(\ref{width}) may be rewritten as\cite{Carena}
\begin{eqnarray}
\Gamma (h \to b \bar{b}) \propto \left ( \frac{\sin \alpha}{\cos
\beta} \right )^2 \left ( \frac{1 - \cot \alpha \cot \beta
\Delta_b}{1 + \Delta_b} \right )^2
\end{eqnarray}
where $\alpha$ is the mixing angle of the two CP-even Higgs boson
obtained by diagonalizing the corresponding mass matrix
${\cal{M}}_H^2$, and $\Delta_b$ is given by
\begin{eqnarray}
\Delta_b&=&\Delta_b^{SQCD} + \Delta_b^{SEW} \nonumber \\
&=& \mu \tan \beta \left ( \frac{2\alpha_s m_{\tilde{g}}}{3\pi}
                        I(m_{\tilde{b}_1},m_{\tilde{b}_2},m_{\tilde{g}})  + \frac{Y_t^2 A_t}{16\pi^2}
                     I(m_{\tilde{t}_1},m_{\tilde{t}_2}, \mu ) +
                     \cdots \right ), \nonumber
\end{eqnarray}
with the function $I$ defined by
\begin{equation}
  I(a,b,c)=\frac{1}{(a^2-b^2)(b^2-c^2)(a^2-c^2)}
  \left(a^2b^2\log{\frac{a^2}{b^2}}
        +b^2c^2\log{\frac{b^2}{c^2}}
        +c^2a^2\log{\frac{c^2}{a^2}}\right)\,.
\end{equation}
Given \begin{eqnarray} {\cal{M}}_H^2 = \left ( \begin{array}{cc}
m_A^2 \sin^2 \beta + m_Z^2 \cos^2 \beta + \Delta_{11} & - (m_A^2 +
m_Z^2 ) \sin \beta \cos \beta + \Delta_{12}
\\ - (m_A^2 + m_Z^2 ) \sin \beta \cos \beta + \Delta_{12} & m_A^2 \cos^2 \beta
+ m_Z^2 \sin^2 \beta + \Delta_{22} \end{array} \right
), \label{CP-even Mass}
\end{eqnarray}
where $\Delta_{ij}$ ($i,j=1,2$) denote the important radiative
corrections with their leading contributions proportional to
$\frac{m_t^4}{m_W^2} \ln \frac{m_{\tilde{t}_1}
m_{\tilde{t}_2}}{m_t^2}$, one can numerically check that without
$\Delta_{ij}$, $\sin\alpha/cos\beta $ is always larger than unity
for $\tan \beta > 7$ as required by muon anomalous momentum. So to
suppress $\sin \alpha$ or equivalently the off-diagonal entry of the
mass matrix the radiative correction must be present, and a positive
large $\Delta_{12}$ along with a light CP-odd Higgs boson is
efficiency in doing this. Meanwhile, given $\cot \alpha \cot \beta
\sim 1$, $\Delta_b$ must be around unity to satisfy $ \cot \alpha
\cot \beta \Delta_b \to 1$, which requires large $\mu \tan \beta$.
In summary, in order to suppress $\Gamma (h_i \to b \bar{b})$, light
$A$ as well as large $\mu \tan \beta$ is favored for given sparticle
spectrum. We note what we are discussing is actually the so-called
`small $\alpha_{\rm eff}$ scenario' of the MSSM \cite{Carena}.

From Eqs.(\ref{definition}-\ref{CP-even Mass}) one can infer that,
in the heavy sparticle limit, the effective $h \gamma \gamma$ and $h
g g$ couplings approach to their SM predictions and $R^{SUSY}$ is
determined by $\Gamma (h_i \to b \bar{b})$ or more generally by the
total width $\Gamma_{tot}(h)$; while in a general case, the
contribution from the sparticle-loops to the couplings may interfere
constructively or destructively with its corresponding SM
contribution, and the size $R^{SUSY}$ then depends on the
competition of $\Gamma ( h \to g g ) \Gamma ( h \to \gamma \gamma )$
with $\Gamma_{tot}(h)$. We checked that this conclusion is also
applicable to the NMSSM and the nMSSM.

\subsection{Results for the MSSM in a general scenario}

\begin{figure}[tbh]
\includegraphics[width=7.cm]{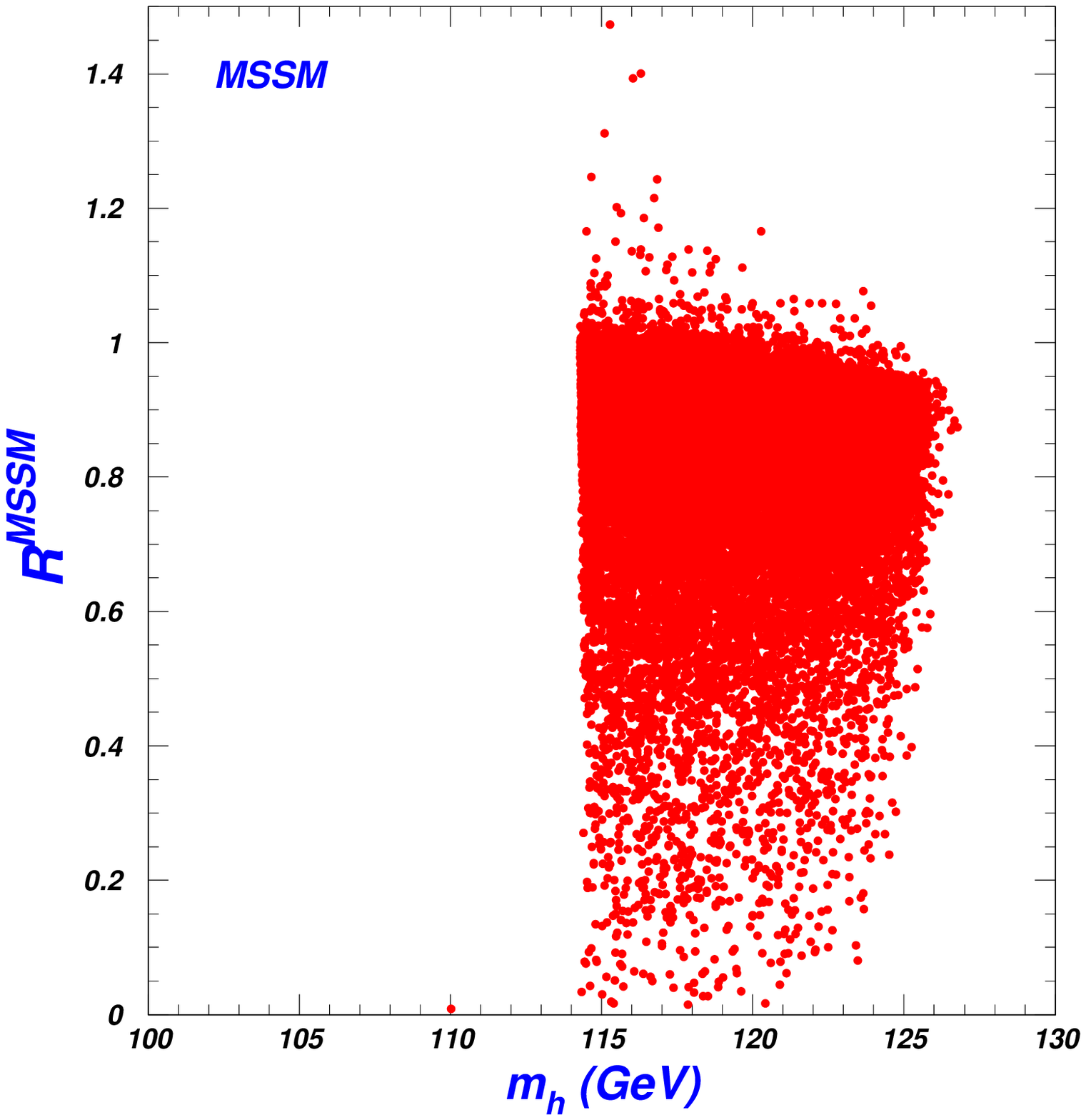}\hspace{0.5cm}
\includegraphics[width=7.cm]{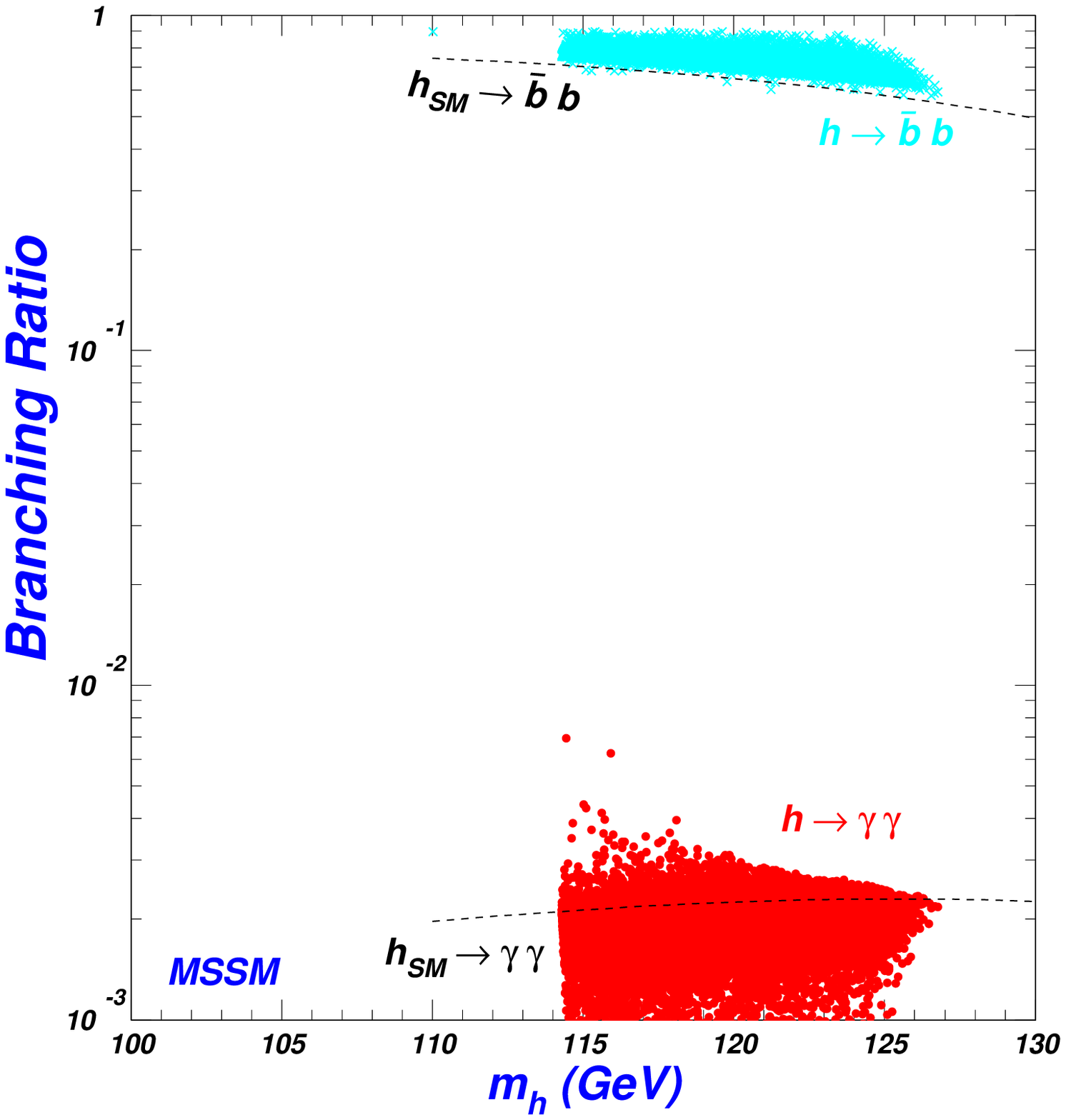}
\vspace*{-0.5cm}
\caption{The scatter plots of the surviving samples, showing the
di-photon signal ratio $R^{MSSM}$ defined in Eq.(\ref{definition})
and the Higgs decay branching ratio versus the mass of
the SM-like Higgs boson.}
\label{fig1}
\end{figure}

To study $R^{SUSY}$ quantatively we scan over the MSSM parameters
under the constraints (1-7) and calculate the di-photon signal rate
for the samples surviving the constraints. Since the first two
generation squarks have little effects on the di-photon signal rate,
in our scan we fix their soft parameters at $1~{\rm TeV}$. As for
sleptons, since it only affects significantly the muon anomalous
magnetic moment $a_\mu$, which can in turn limit the important
parameter $\tan \beta$, we assume all soft parameters in slepton
sector to take a common value $m_{\tilde{l}}$ and treat
$m_{\tilde{l}}$ as a free parameter. For simplicity, we also assume
the grand unification relation for the gaugino masses, $3
M_1/5\alpha_1=M_2/\alpha_2=M_3/\alpha_3$ with $\alpha_i$ being the
fine structure constants of the different gauge groups. Our scan
regions are
\begin{eqnarray}
&&1\leq\tan\beta \leq60,
~~90{\rm ~GeV}\leq m_A\leq 1 {\rm ~TeV}, \nonumber\\
&&200{\rm ~GeV}\leq  M_{\rm SUSY} (=M_{Q_3}=M_{U_3}=M_{D_3})
\leq 1 {\rm ~TeV},\nonumber\\
&&-3{\rm ~TeV}\leq A_{t,b}\leq 3 {\rm ~TeV},
~~100{\rm ~GeV}\leq \mu, M_2, m_{\tilde{l}}\leq 1 {\rm ~TeV} .
\end{eqnarray}
In Fig.\ref{fig1} we display the surviving samples, showing the
di-photon signal ratio $R^{MSSM}$ defined in Eq.(\ref{definition})
and the Higgs decay branching ratio versus the mass of the SM-like
Higgs boson.  This figure shows that in the MSSM there exist some
points where $R$ is enhanced by a factor up to 1.5. Such an
enhancement is mainly due to the suppression of the total width of
$h$, or equivalently the enhancement of Br($h\to \gamma\gamma$),
which is shown in the right frame of Fig.\ref{fig1}. Note that we
required $\mu < 1 {\rm ~TeV}$ in our scan. If we relax $\mu  < 2
{\rm ~TeV}$ in the scan, we find that $R^{MSSM}$ can be as large as
4. We checked that those samples giving $R>1$ actually correspond to
the `small $\alpha_{\rm eff}$ scenario' discussed in \cite{Carena},
which is characterized by a large $\mu \tan \beta$ and $|\sin
\alpha_{eff}/\cos \beta| \leq 1$.

Fig.\ref{fig1} also shows that for most of the samples, the rate of
the di-photon signal is suppressed relative to its SM prediction.
These samples are usually characterized by an enhanced $h\bar{b}b$
coupling and a reduced $h g g$ coupling (the change of the $h\gamma
\gamma$ coupling is usually negligible). We checked that for
$R^{SUSY} > 0.6 $ the effect of the reduced $h g g$ coupling may be
dominant, while for $R^{SUSY} < 0.5 $ the effect of the enhanced
$h\bar{b}b$ coupling is always dominant. We emphasize that for the
samples with $R^{SUSY} < 0.5 $, $A$ must be relatively light ($m_A <
300 {\rm GeV}$) to ensure that the properties of $h$ significantly
deviate from the SM Higgs boson \cite{Gunion}.

We note that current experiments can not rule out a light $A$ with
$110 {\rm ~GeV} < m_A < 140 {\rm ~GeV}$ in the MSSM \cite{Cao-MSSM}.
In this case, both $A$ and $H$ (the heavier CP-even Higgs boson)
give rise to the di-photon signals similar to the SM-like Higgs
boson $h$. However, the rates of these signals from $A$ and $H$ can
not be large. This is because for $110 {\rm ~GeV} < m_A < 140 {\rm
~GeV}$, $\tan \beta$ must be larger than $7$ as required by the
constraints (particularly by $a_\mu$) \cite{Cao-MSSM}, which implies
$\cos \alpha > 0.8 $ from the tree-level relation $\tan 2 \alpha =
\tan 2 \beta \frac{m_A^2+ m_Z^2}{m_A^2 - m_Z^2}$. Since the $A
\bar{b}b$ and $H\bar{b} b$ couplings are proportional to $\tan
\beta$ and $\cos \alpha/\cos \beta$ respectively, the branching
ratios of $A,H \to \gamma \gamma$ are suppressed and so are their
induced di-photon signals at the LHC \cite{Gunion-fourth}.

\vspace{-0.5cm}

\subsection{Results for different models in the $m_h^{max}$ scenario}
Since the NMSSM and the nMSSM have more free parameters than the
MSSM, it is difficult to perform a general analysis of the signal
rate. However, considering our aim is to show the differences of
these three models, we examine the signal in the so-called
$m_h^{max}$ scenario described in Sec. \ref{sec-a}. In this
scenario, under the constraints (1-7) we scan over the following
parameter ranges:
\begin{eqnarray}
&&90{\rm ~GeV}\leq m_A\leq 1 {\rm ~TeV},
1\leq\tan\beta \leq60, ~100{\rm ~GeV}\leq \mu, M_2,
m_{\tilde{l}} \leq 1 {\rm ~TeV}, \nonumber\\
&& 100{\rm ~GeV}\leq  M_{\rm SUSY} (=M_{Q_3}=M_{U_3}=M_{D_3}) \leq 1
{\rm ~TeV} ,
\end{eqnarray}
for the MSSM,
\begin{eqnarray}
&& 0<\lambda, \kappa \leq 0.7, ~90{\rm ~GeV}\leq m_A\leq 1 {\rm ~TeV},
\nonumber\\
&& 100{\rm ~GeV}\leq  M_{\rm SUSY} (=M_{Q_3}=M_{U_3}=M_{D_3})
\leq 1 {\rm ~TeV} , \nonumber\\
&& 1\leq\tan\beta \leq 60, ~|A_{\kappa}|\leq 1{\rm ~TeV}, ~~100{\rm
~GeV}\leq \mu, M_2, m_{\tilde{l}}\leq 1 {\rm ~TeV},
\end{eqnarray}
for the NMSSM, and
\begin{eqnarray}
&&0.01\leq \lambda\leq 0.7,~~
100{\rm ~GeV}\leq m_A, \mu, M_2\leq 1000 {\rm ~GeV},\nonumber\\
&& 100{\rm ~GeV}\leq  M_{\rm SUSY} (=M_{Q_3}=M_{U_3}=M_{D_3}) \leq 1
{\rm ~TeV},  \nonumber \\ &&1\leq\tan\beta \leq 60, ~~-1{\rm
~TeV}\leq A_{\lambda}\leq 1{\rm ~TeV},
 ~~0\leq\tilde m_S\leq 200{\rm ~GeV}, \label{nMSSM-scan}
\end{eqnarray}
for the nMSSM with the soft parameters to be 100 GeV for the
($\tilde \nu_{\mu},\tilde \mu$) sector in order to satisfy the
$a_\mu$ constraint\cite{cao-xnmssm}. For other insensitive
parameters we adopt the same assumption as in the last section.

\begin{figure}[tbh]
\centering
\includegraphics[width=5.5cm]{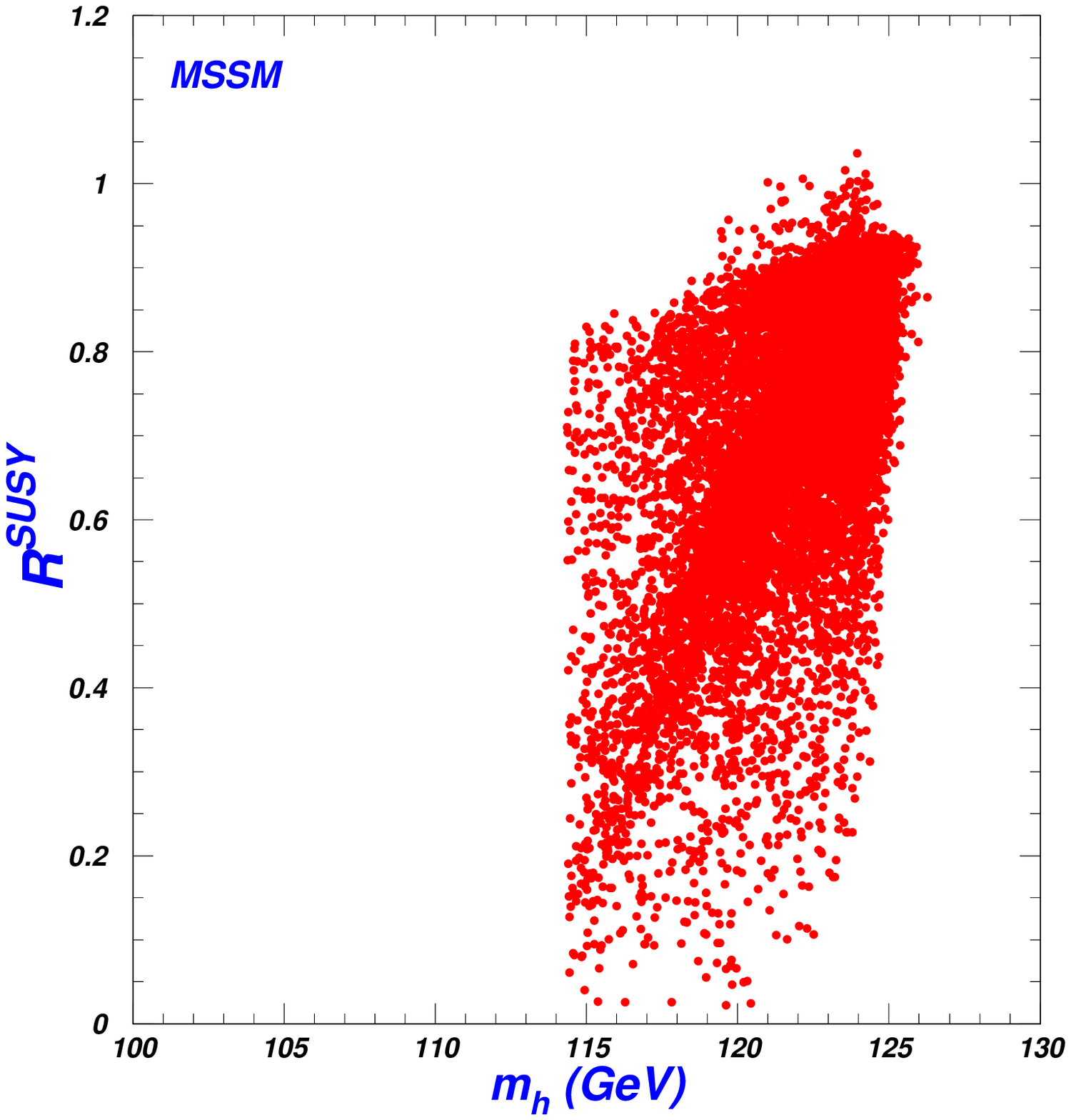}\hspace{-0.2cm}
\includegraphics[width=5.5cm]{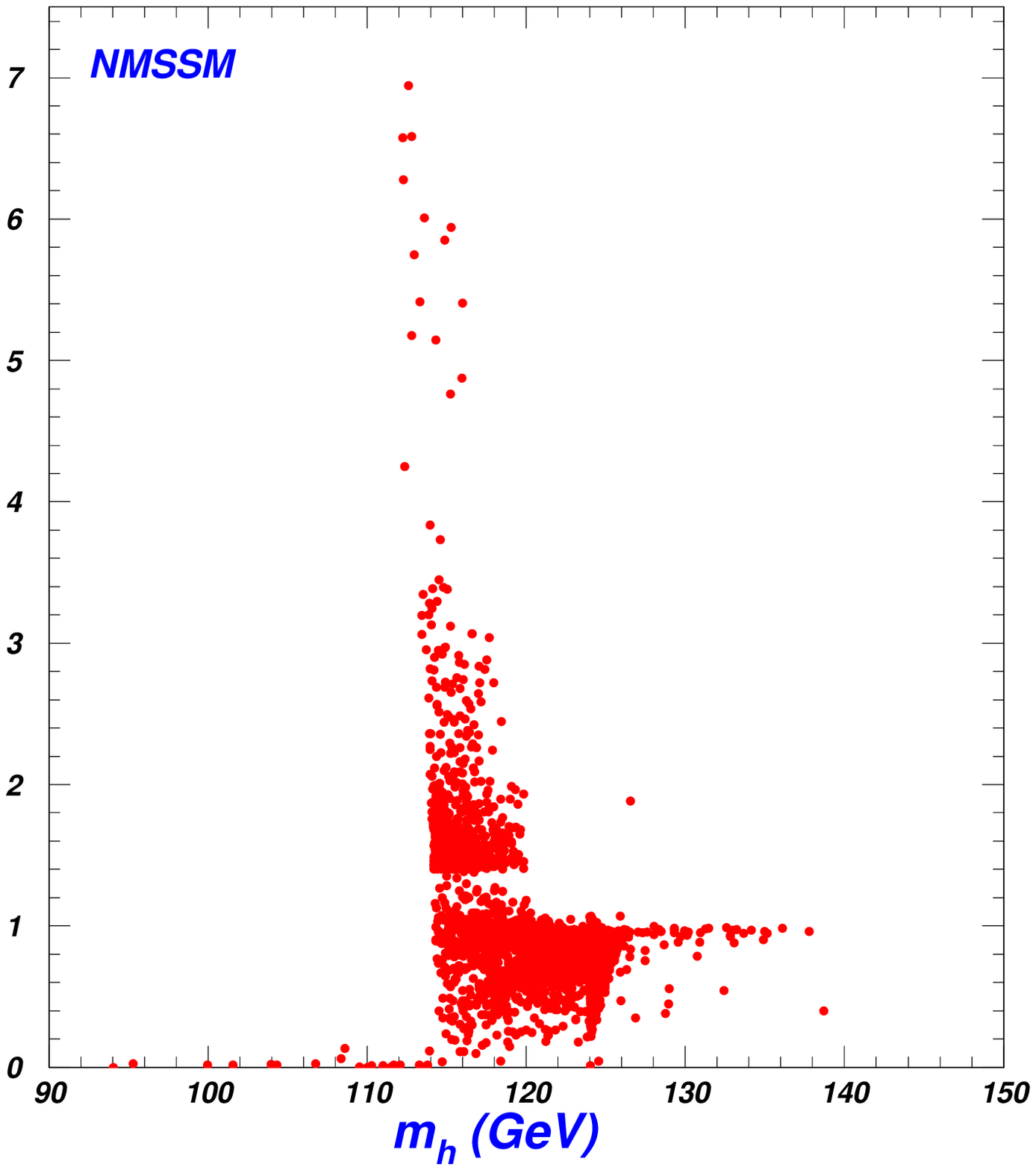}\hspace{-0.2cm}
\includegraphics[width=5.5cm]{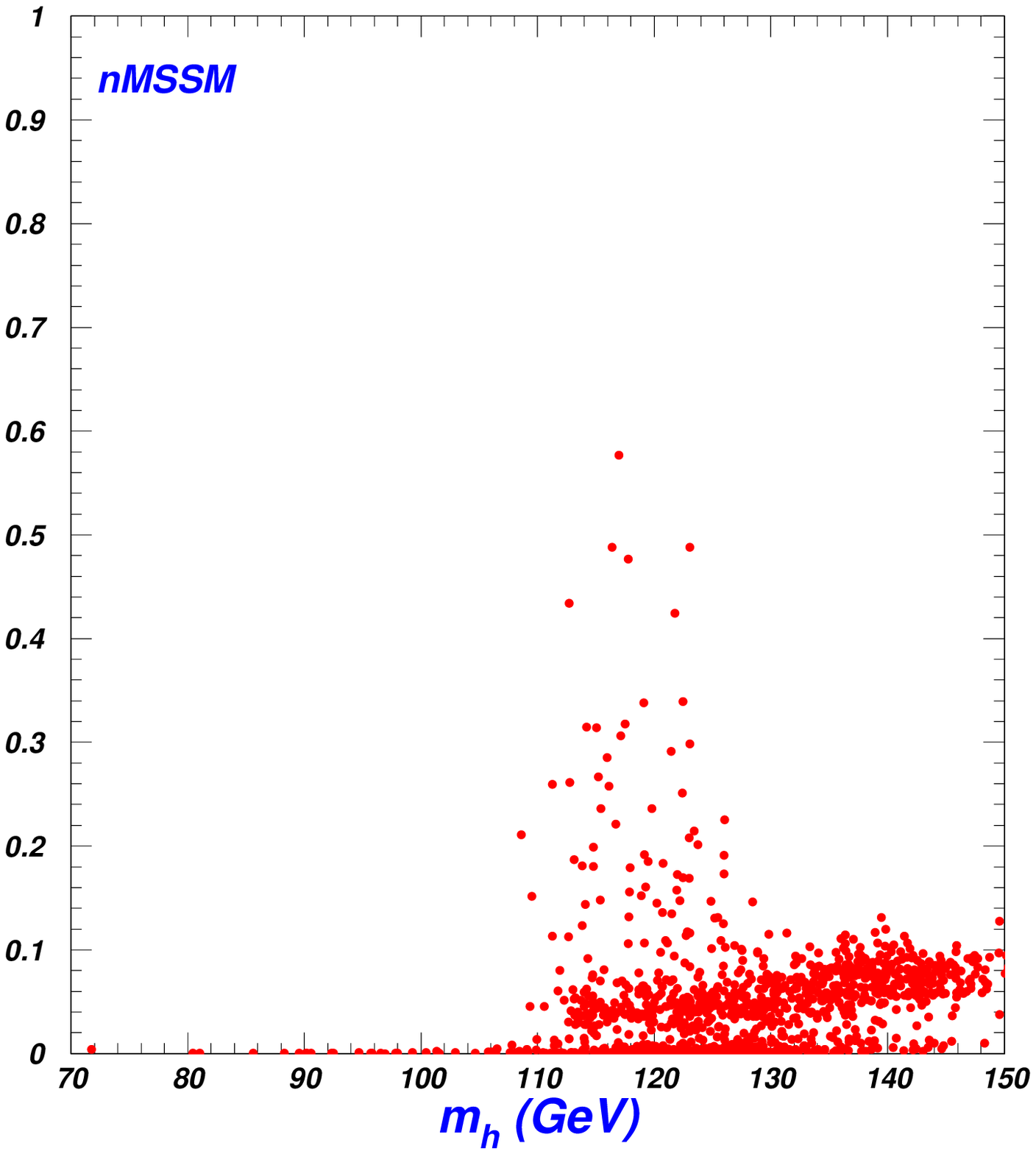}
\vspace*{-0.5cm}
\caption{The scatter plots of the surviving samples in the
$m_h^{max}$ scenario of the MSSM, NMSSM and nMSSM, showing the
di-photon signal ratio defined in Eq.(\ref{definition})
versus the mass of the SM-like Higgs boson.}
\label{fig2}
\end{figure}

\begin{figure}[htb]
\centering
\includegraphics[width=7.5cm]{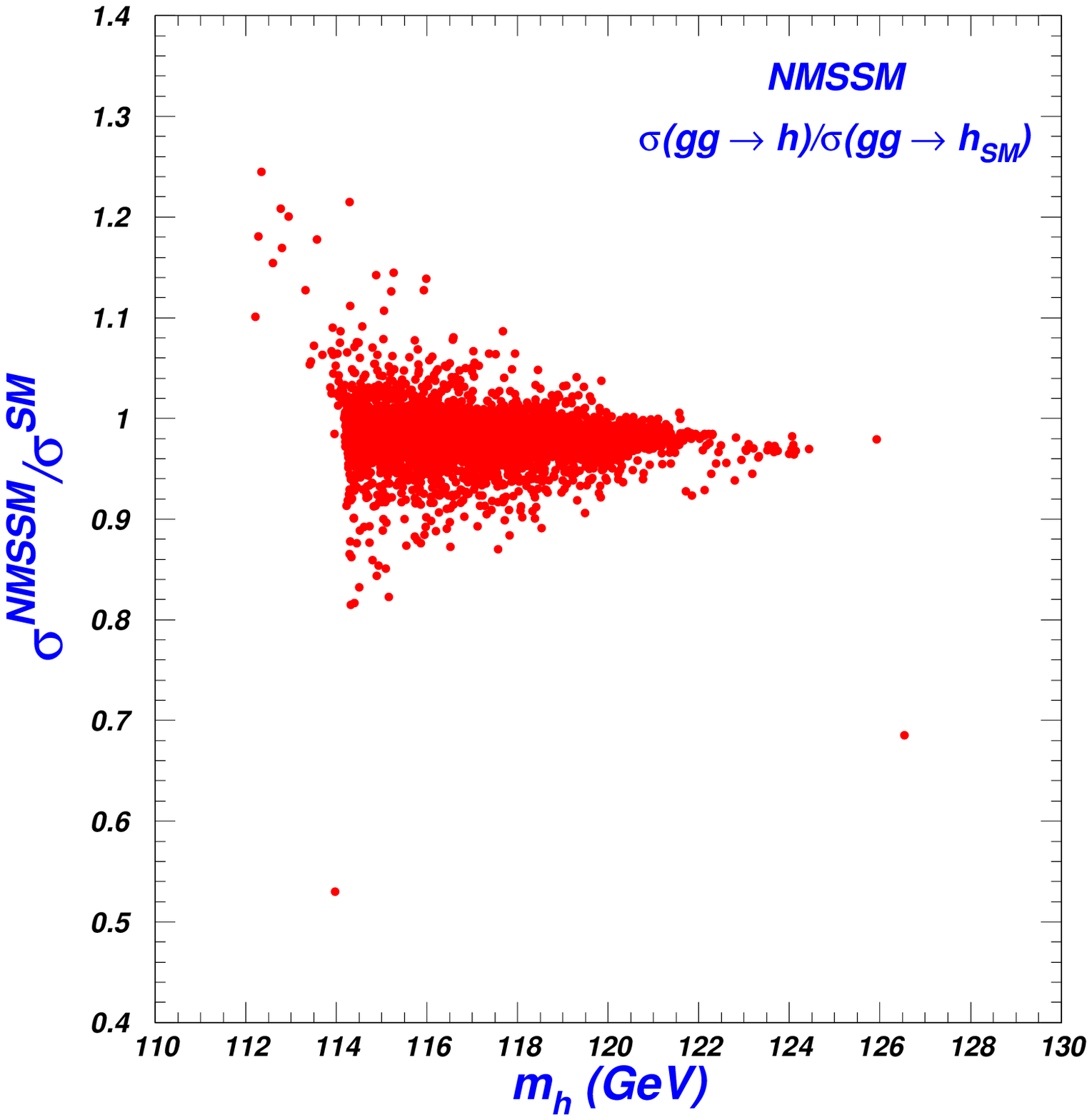}\hspace{0.5cm}
\includegraphics[width=7.5cm]{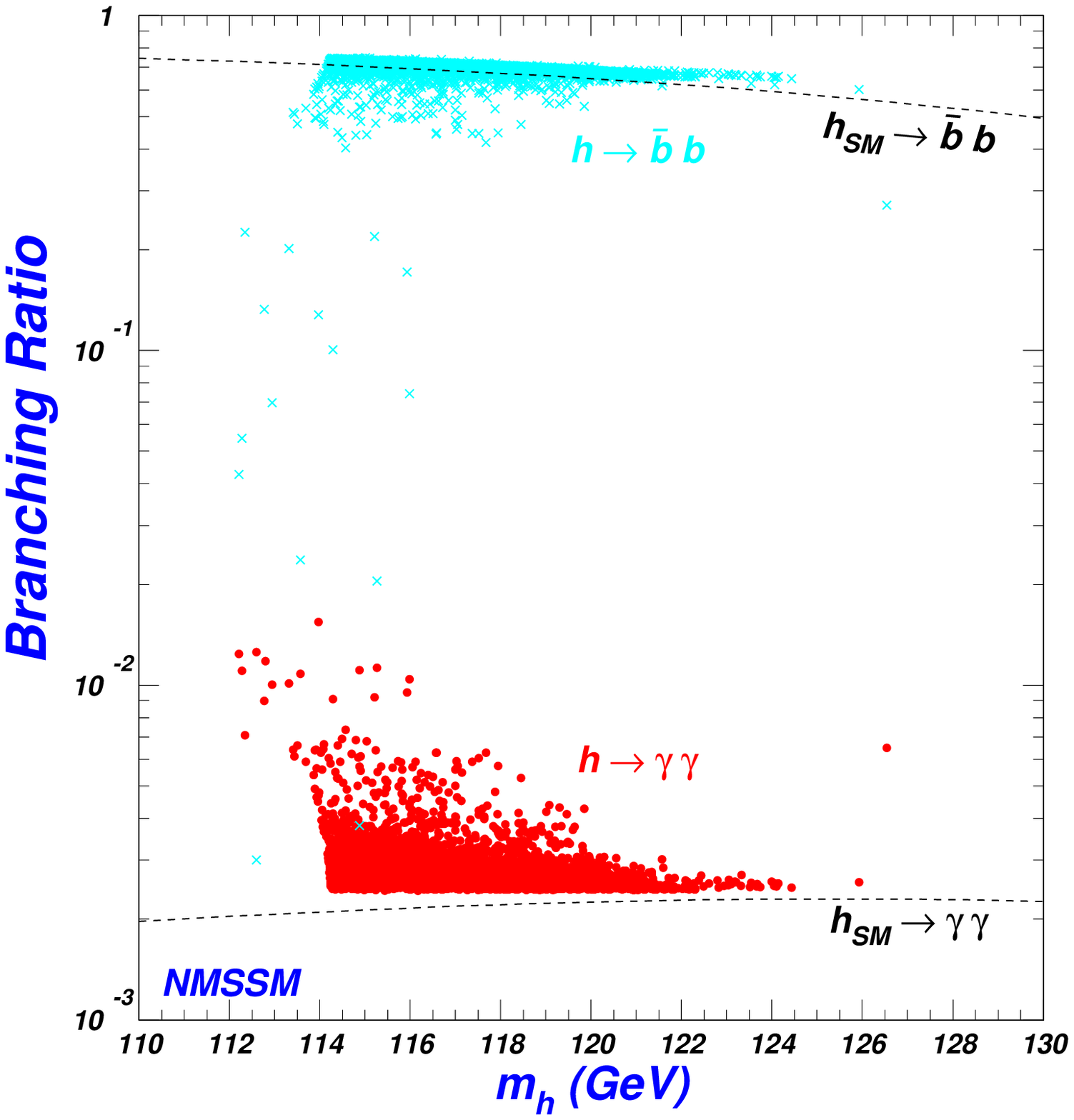}
\vspace*{-0.5cm}
\caption{Same as Fig.2, but projected on different planes for the
NMSSM. Here only the samples satisfying $R^{\rm NMSSM}>1$ are plotted.}
\label{fig3}
\end{figure}

\begin{figure}[htb]
\centering
\includegraphics[width=10cm]{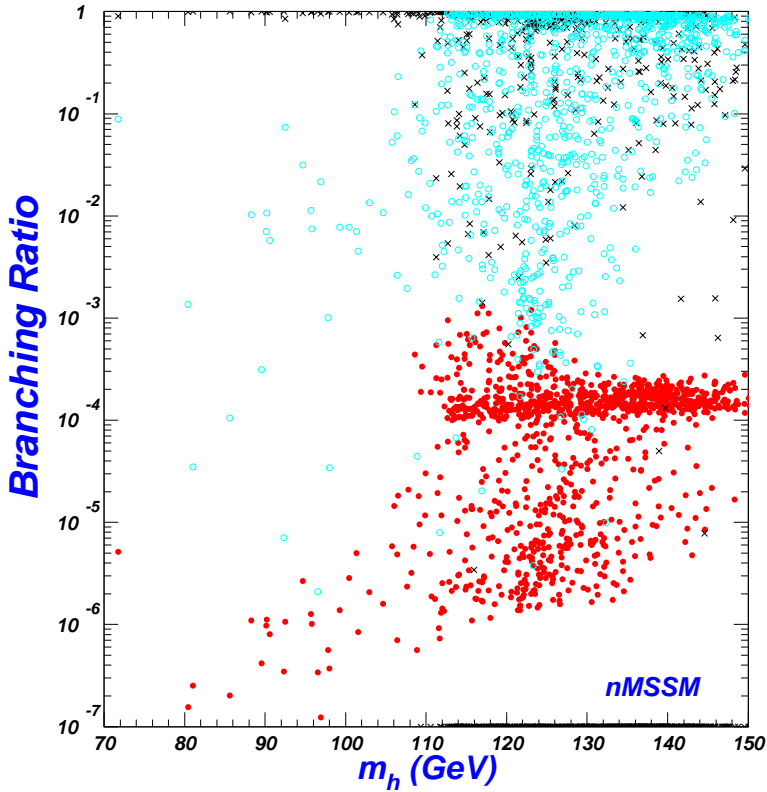}
\vspace*{-0.5cm}
\caption{Same as Fig.2, but showing the branching ratios of the
SM-like Higgs decay to $\gamma\gamma$ (`$\bullet$', red), to $a a$
(`$\times$', black), and to $\tilde\chi_1^0 \tilde\chi_1^0$
(`$\circ$', sky blue) in the nMSSM.} \label{fig4}
\end{figure}

In Fig.\ref{fig2} we show the di-photon signal rates in the
$m_h^{max}$ scenario for three models. This figure shows that in the
nMSSM the signal is always suppressed. In the MSSM the signal is
mostly suppressed, but in a tiny part of the parameter space the
signal can be slightly enhanced. In the NMSSM, however, the signal
can be enhanced in a sizable part of the parameter space (the
enhancement factor can be as large as 7). In order to figure out the
reason for such a large enhancement in the NMSSM, we concentrate on
the samples with $R >1 $  and study the ratio
$\sigma^{NMSSM}/\sigma^{SM} (p p \to h)$ and $Br(h \to b \bar{b},
\gamma \gamma)$. Our results are shown in Fig.\ref{fig3}, which
indicates that the production rate can be enhanced maximally by a
factor of 1.25,  while  $Br(h \to \gamma \gamma)$ can be enhanced to
$2 \times 10^{-2}$ once $Br(h \to b \bar{b})$ is suppressed to
several percent. This conclusion justifies our previous analysis
about the mechanisms to enhance the signal.

Quite surprisingly, we found that in the NMSSM the samples with
$R\gg 1$ are unnecessarily accompanied by a large $\mu$. The
fundamental reason is that in SUSY  the $h\bar{b}b$ coupling is
determined by the $H_d$ component of $h$, and in the NMSSM, due to
the presence of singlet field component in $h$, the $h\bar{b}b$
coupling can be suppressed more efficiently than in the MSSM. We
also noticed that once $Br(h \to b \bar{b})$ is suppressed, $Br(h
\to V V^\ast)$ ($V = W, Z$) may also get enhanced, which should be
limited by the combined search for Higgs boson at the Tevatron
\cite{Tevatron-2010}. We checked that for $Br(h \to \gamma
\gamma)\sim 10^{-2}$, $Br(h \to V V^\ast)$ can be enhanced by a
factor of 4 relative to its SM prediction.

For the samples with a suppressed di-photon rate in the NMSSM and
the nMSSM, we find that $\Gamma_{tot}(h)$ is usually enhanced (due
to the enhanced $h \bar{b} b$ and/or the open-up of new decay modes
) and the $h g g$ coupling is reduced. We checked that for $R^{SUSY}
< 0.5$ the former effect is  dominant. We note that in the nMSSM
$R^{SUSY}$ is usually small, which is mainly due to the open-up of
new decay modes of $h$, such as  $h \to \tilde{\chi}_1^0
\tilde{\chi}_i^0$ ($i=1,2$) or $ h \to a a$ with their rates shown
in Fig.\ref{fig4}. We emphasize that this feature comes from the
peculiarity of $\tilde{\chi}_1^0$ in the nMSSM (see
Eq.(\ref{mass-exp})) and should keep valid regardless our choice of
the soft parameters in the squark sector. We numerically checked
this point by a more general scan than Eq.(\ref{nMSSM-scan}). We
also note that for nearly all the samples in the NMSSM with $m_h
> 120 {\rm GeV}$ we have $R^{SUSY} < 1$, and for all the samples in the
nMSSM with $m_h > 125 {\rm GeV}$ we have $R^{SUSY} < 0.14$. We owe
this to the constraints we considered, which severely constrained
the enhancement of the branching ratio of $h \to \gamma \gamma$
(see Fig.\ref{fig3} and Fig.\ref{fig4}).

We also studied the di-photon signal rate in the `no-mixing'
scenario defined as $A_t=A_b$ and $X_t =0$. However, we found it is
difficult for this scenario to satisfy the constraints if $M_{SUSY}
< 1 {\rm ~TeV}$, especially we did not find any surviving samples
for the MSSM. Since the di-photon signal for the surviving samples
in the NMSSM and the nMSSM do not exhibit new characteristics, we do
not present the results here.

So, we see that in low energy SUSY, depending on the models,
the di-photon signal rate at the LHC may be significantly
suppressed or enhanced relative to the SM prediction.
With $2 fb^{-1}$ integrated luminosity at the running LHC,
the di-photon signal can allow for a test
of the SM and a probe of the low energy SUSY models.
For example, if the di-photon signal rate is found to be not smaller
than the SM prediction, then the nMSSM will be immediately excluded
(note that in this case the universal extra dimension and the little
Higgs theory will also be ruled out because they suppressed the diphoton
signal rate \cite{Yuan,wang}).

\section{Conclusion}
We focused on the di-photon Higgs signal $gg\to h\to\gamma\gamma$
for the SM-like Higgs boson at the LHC and performed a comparative
study for three SUSY models: the MSSM, NMSSM and nMSSM.
Considering various collider and
cosmological constraints, we scanned over the parameter space
and obtained the following observation in the allowed parameter space:
(i) In the nMSSM the signal rate is always suppressed;
(ii) In the MSSM the signal rate is suppressed in most cases, but in
a tiny corner of the parameter space it can be enhanced (maximally
by a factor of 2);
(iii) In the NMSSM the signal rate can be suppressed or enhanced
depending on the parameter space, and the enhancement factor
can be as large as 7.

{\rm Note added:~} After we finished the manuscript, we noticed a
preliminary result from the ATLAS collaboration \cite{latest}, which
excluded $R \geq 4.2$ ($R$ is defined in Eq.\ref{definition}) for
$m_h \simeq 115 {\rm GeV}$. This means that in the middle panel of
Fig.2 the samples above $R\simeq 4.2$ for the NMSSM will be
excluded.

\section*{Acknowledgement}
 JMY thanks JSPS for the invitation
fellowship (S-11028) and the particle physics group of
Tohoku University for their hospitality.
This work was supported in part by HASTIT under grant No.
2009HASTIT004, by the National Natural Science Foundation of China
(NNSFC) under grant Nos. 10821504, 10725526, 10775039,
11075045 and by the Project of Knowledge Innovation Program (PKIP)
of Chinese Academy of Sciences under grant No. KJCX2.YW.W10.


\begin{thebibliography}{99}

\bibitem{h-mass-ew} M. Goebel (for the Gfitter Group), arXiv:0905.2488.
\bibitem{Barate:2003sz}
  R.~Barate {\it et al.},
  Phys.\ Lett.\  B {\bf 565}, 61 (2003).

\bibitem{Tevatron-2010}    [CDF and D0 Collaboration],
  arXiv:1007.4587 [hep-ex].

\bibitem{LHCHiggsCross}
  LHC Higgs Cross Section Working Group {\it et al.},
  arXiv:1101.0593 [hep-ph].

\bibitem{LHC}
  I.~T.~f.~Collaboration,
  arXiv:1012.0694 [hep-ex].

\bibitem{Yuan}  K.~Hsieh and C.~P.~Yuan,
  Phys.\ Rev.\  D {\bf 78}, 053006 (2008).

\bibitem{diphoton1}
  I.~Low and S.~Shalgar,
  JHEP {\bf 0904}, 091 (2009).

\bibitem{diphoton2} S.~Moretti and S.~Munir,
  Eur.\ Phys.\ J.\  C {\bf 47}, 791 (2006);
  U.~Ellwanger,
  arXiv:1012.1201.

\bibitem{diphoton3}  M.~Almarashi and S.~Moretti,
  arXiv:1011.6547 [hep-ph].

\bibitem{NMSSM1}
For a review, see, e.g., U.~Ellwanger, C.~Hugonie and A.~M.~Teixeira,
  Phys.\ Rept.\  {\bf 496}, 1 (2010);

\bibitem{NMSSM2} For phenomenological studies, see, e.g.,
J.~R.~Ellis  {\it et al.}, \PRD39, 844  (1989);
M.~Drees, Int. J. Mod. Phys. A{\bf 4}, 3635  (1989);
S.~F.~King, P.~L.~White, \PRD52, 4183 (1995);
B. Ananthanarayan, P.N. Pandita, \PLB353, 70  (1995);
B. A. Dobrescu, K. T. Matchev, \JHEP0009, 031  (2000);
V. Barger {\it et al.}, \PRD73, 115010  (2006);
R.~Dermisek, J.~F.~Gunion, \PRL95, 041801 (2005);
G.~Hiller, \PRD70, 034018 (2004);
F.~Domingo, U.~Ellwanger, \JHEP0712, 090  (2007);
Z.~Heng {\it et al.}, \PRD77, 095012  (2008);
R. N. Hodgkinson,  A. Pilaftsis, \PRD76, 015007  (2007);
W. Wang {\it et al.}, \PLB680, 167  (2009);
 J.~Cao {\it et al.}, \JHEP0812, 006 (2008);  \PRD78, 115001 (2008);
 J. M. Yang, arXiv:1102.4942.


\bibitem{xnMSSM} C.~Panagiotakopoulos, K.~Tamvakis,
   \PLB446, 224 (1999); \PLB469, 145 (1999);
  C.~Panagiotakopoulos, A. Pilaftsis, \PRD63, 055003 (2001);
  A.~Dedes, {\it et al.}, \PRD63, 055009 (2001);
  A.~Menon, {\it et al.}, \PRD70, 035005 (2004);
  V.~Barger, {\it et al.}, \PLB630, 85 (2005).
  C.~Balazs, {\it et al.}, \JHEP0706, 066 (2007).

\bibitem{cao-xnmssm}  J.~Cao, H.~E.~Logan and J.~M.~Yang,
  Phys.\ Rev.\  D {\bf 79}, 091701 (2009);
                    J.~Cao, Z.~Heng and J.~M.~Yang,
  JHEP {\bf 1011}, 110 (2010).

\bibitem{lsp-mass}  S. Hesselbach, {\it et al.}, arXiv:0810.0511v2 [hep-ph].

\bibitem{Djouadi:1998az}
  A.~Djouadi,
  Phys.\ Lett.\  B {\bf 435}, 101 (1998)
  [arXiv:hep-ph/9806315].

\bibitem{PDG2010}
  K.~Nakamura {\it et al.}  [Particle Data Group],
  J.\ Phys.\ G {\bf 37}, 075021 (2010).

\bibitem{Abdallah}
  J.~Abdallah {\it et al.},
  Eur.\ Phys.\ J.\  C {\bf 31}, 421 (2004);
 G.~Abbiendi {\it et al.},
  Eur.\ Phys.\ J.\  C {\bf 35}, 1 (2004).

\bibitem{Altarelli}
  G.~Altarelli and R.~Barbieri, \PLB253, 161 (1991);
  M. E. Peskin, T. Takeuchi, \PRD46, 381 (1992).

\bibitem{Abazov}
  V.~M.~Abazov {\it et al.}  [D0 Collaboration],
  Phys.\ Rev.\ Lett.\  {\bf 103}, 061801 (2009).

\bibitem{Davier}
  M.~Davier {\it et al.}, \EPJC66, 1 (2010).

\bibitem{Dunkley}
  J.~Dunkley {\it et al.}  [WMAP Collaboration],
  Astrophys.\ J.\ Suppl.\  {\bf 180}, 306 (2009).

\bibitem{NMSSMTools}  U.~Ellwanger, J.~F.~Gunion and C.~Hugonie,
  JHEP {\bf 0502}, 066 (2005);
                   U.~Ellwanger and C.~Hugonie,
  Comput.\ Phys.\ Commun.\  {\bf 175}, 290 (2006).


\bibitem{Carena}
  M.~S.~Carena, S.~Heinemeyer, C.~E.~M.~Wagner and G.~Weiglein,
  Eur.\ Phys.\ J.\  C {\bf 26}, 601 (2003).


\bibitem{STU} J.~Cao {\it et al.}, \JHEP1007, 044 (2010).

\bibitem{Carena-eff}  M.~S.~Carena, D.~Garcia, U.~Nierste and C.~E.~M.~Wagner,
  Nucl.\ Phys.\  B {\bf 577}, 88 (2000).

\bibitem{Gunion}
  J.~F.~Gunion and H.~E.~Haber,
  Phys.\ Rev.\  D {\bf 67}, 075019 (2003).

\bibitem{Cao-MSSM}  J.~Cao {\it et al.}, \PRD82, 051701 (2010).

\bibitem{Gunion-fourth}  J.~F.~Gunion,
  arXiv:1105.3965 [hep-ph].

\bibitem{wang}  L.~Wang and J.~M.~Yang,
  Phys.\ Rev.\  D {\bf 79}, 055013 (2009).

\bibitem{latest} The ATLAS collaboration, ATLAS-CONF-2011-085.

\end{thebibliography}
\end{document}